# Tracking the evolution of magmas from heterogeneous mantle sources to eruption

**A. Mallik[1†], S. Lambart[2†], and E.J. Chin[3†]**

[1]Department of Geosciences, University of Rhode Island.

[2]Department of Geology and Geophysics, University of Utah.

[3]Scripps Institution of Oceanography, University of California, San Diego.

Corresponding author: Ananya Mallik (ananya_mallik@uri.edu)

†Equally contributing authors

**Key Points:**

- Effects of source heterogeneity and intracrustal differentiation on magma chemistry is explored
- Discussion is based on experimental studies and compositions of about 60,000 natural volcanic rocks, melt inclusions and cumulates
- Magmatism in mid-ocean ridges, ocean islands and arcs are considered.

**Abstract**

This contribution reviews the effects of source heterogeneities, melt-rock reactions and intracrustal differentiation on magma chemistry across mid-ocean ridges, intraplate settings and subduction zones using experimental studies and natural data. We compare melting behaviors of pyroxenites and peridotites and their relative contributions to magmas as functions of composition, mantle potential temperatures and lithospheric thickness. We also discuss the fate of chemically distinct melts derived from heterogeneities as they travel through a peridotitic mantle. Using nearly 60,000 natural major element compositions of volcanic rocks, melt inclusions, and crystalline cumulates, we assess broad petrogenetic trends in as large of a global dataset as possible. Consistent with previous studies, major element chemistry of mid-ocean ridge basalts (MORBs) and their cumulates favor a first-order control of intracrustal crystal-liquid segregation, while trace element studies emphasize the role of melt-rock reactions, highlighting the decoupling between the two. Ocean island basalts (OIB) show a larger compositional variability than MORB, partly attributed to large variations of pyroxenite proportions in the mantle source. However, the estimated proportions vary considerably with heterogeneity composition, melting model and thermal structure of the mantle. For arcs, we highlight current views on the role of the downgoing slab into the source of primary arc magmas, and the role of the overriding lithosphere as a magmatic chemical filter and as the repository of voluminous arc cumulates. Our approach of simultaneously looking at a large database of volcanic + deep crustal rocks across diverse tectonic settings underscores the challenge of deciphering the source signal versus intracrustal/lithospheric processes.

**Plain Language Summary**

Volcanic rocks on Earth share a common history – melting in the mantle tens to hundreds of kilometers deep, traveling through the mantle to the crust, and cooling and partial crystallizing in the crust, before finally reaching the surface – and yet, they show an incredible compositional variability. What creates such variability? In this chapter, we discuss the potential causes by looking at the role of the heterogeneities in the mantle and the fate of the magmas along their journey to the surface through the mantle and the crust. The discussion is based on experimental studies and natural compositions of about 60,000 volcanic rocks, melt inclusions (small pools of magma trapped inside the minerals) and cumulates (products of the crystallization of the magmas) from three main geological settings: divergent plate boundaries (mid-ocean ridges), intra tectonic plates (oceanic islands) and convergent plate boundaries (subduction zones).

**1 Introduction**

Magmatism on Earth takes place at three main different tectonic settings: mid-ocean ridges, intraplate settings including ocean islands, and arc magmatism in subduction zones. The upper mantle, mostly composed of peridotites, is the main source of magma on Earth. However, geophysical, geochemical and field observations demonstrate that the mantle is heterogeneous. Many studies have proposed that crustal lithologies introduced in the Earth's mantle by subduction, including sediments and oceanic crust contribute significantly as heterogeneities to the generation of mantle-derived magmas (e.g. Chase, 1981; Helffrich and Wood, 2001; Hirschmann and Stolper, 1996; Hofmann, 1997; Hofmann and White, 1982; Jackson et al., 2007; Lambart, 2017; Lambart et al., 2009, 2016; Mallik and Dasgupta, 2012; Salters and Dick, 2002; Schiano et al., 1997; Sobolev et al., 2005, 2007). The contribution of a lithology during magma genesis is controlled by four main parameters: (1) the fraction of the lithology in the mantle, (2) its solidus temperature, (3) its melt productivity, and (4) the mantle regime (potential temperature and thickness of the lithosphere). In this chapter, we discuss the current state of understanding of how these parameters affect the contribution of heterogeneities in magma generation in the Earth's upper mantle across the three tectonic settings. We also review perspectives on magmatic differentiation, since intracrustal differentiation is the "last mile" of a primary magma's journey from its origin as a partial melt of the mantle to its point of eruption. Disentangling the contributions from differentiation vs. source heterogeneity are therefore non-trivial. To these ends, we review results from experimental studies simulating partial melting under a variety of upper mantle conditions and compositions. Then we compare experimental results with a compilation of nearly 60,000 compositions of natural volcanic rocks, melt inclusions, and plutonic cumulate rocks from ridges, intraplate settings, and volcanic arcs.

**2 Partial melting of a heterogeneous mantle**

*2.1 Contribution of mantle lithologies to magma genesis*

For the purpose of this section, we use the term "pyroxenite" to refer to mafic and ultramafic-rich plutonic rocks that lack sufficient olivine (40%) to be classified as peridotites (Le Maitre et al., 2005). Based on their abundance in orogenic massifs, it is generally assumed that the Earth's upper mantle contains 2 to 5% of pyroxenites (e.g., Bodinier and Godard, 2003). Magma compositions suggest that the mantle source can be significantly enriched locally (e.g., Sobolev et al., 2005) or even be exclusively composed of pyroxenites (e.g., Zhang et al., 2018). Compositions of the lithologies control their solidus temperatures and melt productivities (Hirschmann, 2000; Lambart et al., 2016; Pickering-Witter and Johnston, 2000), while thermal structure of the mantle and the crust control the melt fluxes and initial and final pressures of melting (Langmuir et al., 1992; Sleep, 1990; Tatsumi et al., 1983). The influence of composition on the solidus temperatures of mantle lithologies has been the subject of several experimental and theoretical studies (Hirschmann, 2000; Kogiso and Hirschmann, 2006; Kogiso et al., 2004; Lambart et al., 2016). These studies demonstrate that, at a given pressure, the solidus temperatures of anhydrous and $CO_2$-free lithologies are mostly influenced by the bulk alkali content **(Fig. 1)** and Mg-number (Mg#=molar MgO/[molar MgO+molar $FeO^T$] of the rock.

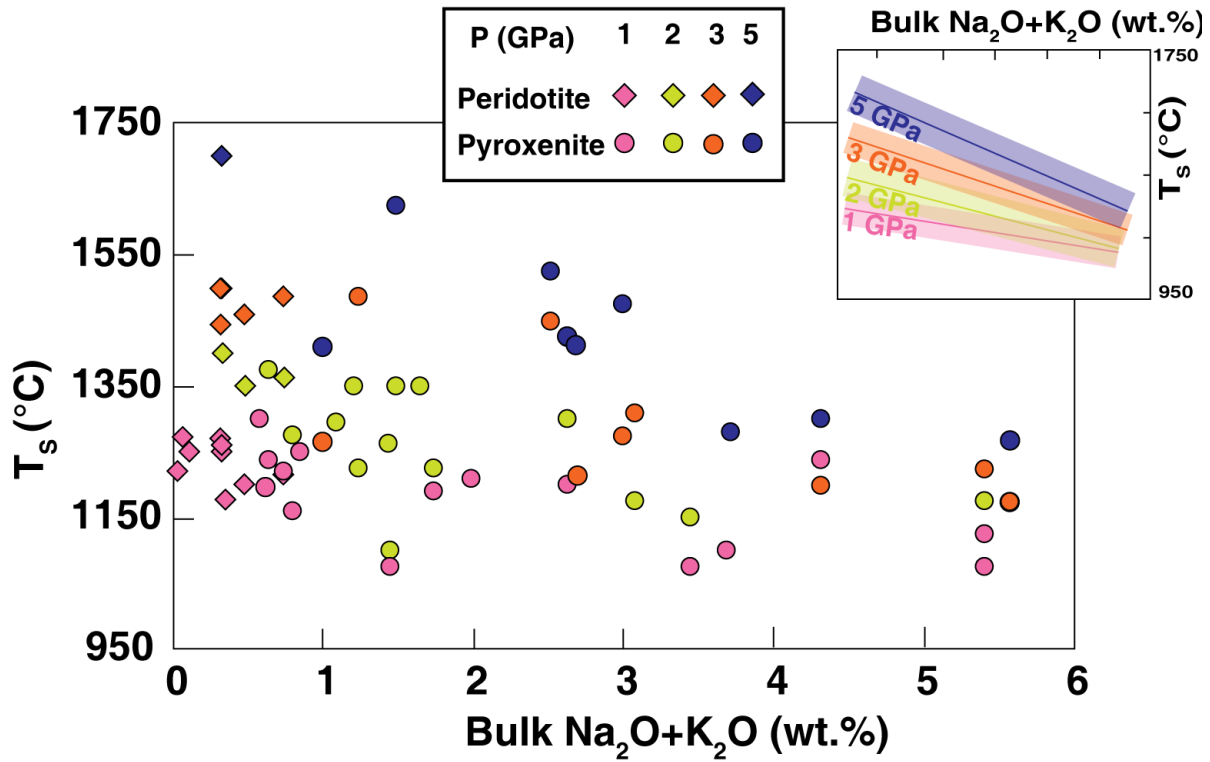

**Figure 1.** Pyroxenite and peridotite solidus temperatures determined experimentally as a function of the bulk alkali content for pressures between 1 and 5 GPa. The inset shows the linear regressions for each pressure and the corresponding error envelopes (1$\sigma$). References: Adam et al., 1992; Baker and Stolper, 1994; Borghini et al., 2017; Hirose and Kushiro, 1993; Hirschmann et al., 2003; Ito and Kennedy, 1974; Kogiso et al., 1998; Kogiso and Hirschmann, 2001,2006; Kogiso et al., 2003; Kornprobst, 1970; Kushiro, 1996; Lambart et al., 2009, 2012; Pertermann and Hirschmann, 2003a; Pickering-Witter and Johnston, 2000; Rosenthal et al., 2014, 2018; Schwab and Johnston, 2001; Spandler et al., 2008, 2010; Thompson, 1974, 1975; Tsuruta and Takahashi, 1998; Tuff et al., 2005; Walter, 1998; Wasylenki et al., 2003; Yasuda et al., 1994

Fewer studies, however, discuss the controls on melt productivity. Melt productivities of mantle lithologies tend to increase with the contribution of clinopyroxene (cpx; Lambart et al., 2013; Pertermann and Hirschmann, 2003a; Pickering-Witter and Johnston, 2000; Spandler et al., 2008) and feldspar (Falloon et al., 2008; Lambart et al., 2009; Spandler et al., 2008) in the melting reaction. On the contrary, the presence of other accessory phases like rutile, quartz, spinel or olivine results in low melt productivities, explained by a low thermodynamic variance of the system (Pertermann and Hirschmann, 2003a). **Figure 2** compares experimental melt fractions as a function of temperature at 1 and 3 GPa for various pyroxenite and peridotite compositions. The higher solidus temperature difference between pyroxenites and peridotites at 3 GPa than at 1 GPa partly reflects sampling bias (most experiments at 3 GPa were performed on alkali-richer and Mg-poorer lithologies than experiments performed at 1 GPa), but is also consistent with the parameterization of Lambart et al. (2016) that shows that the solidus temperature variation in pyroxenites significantly decreases with pressure.

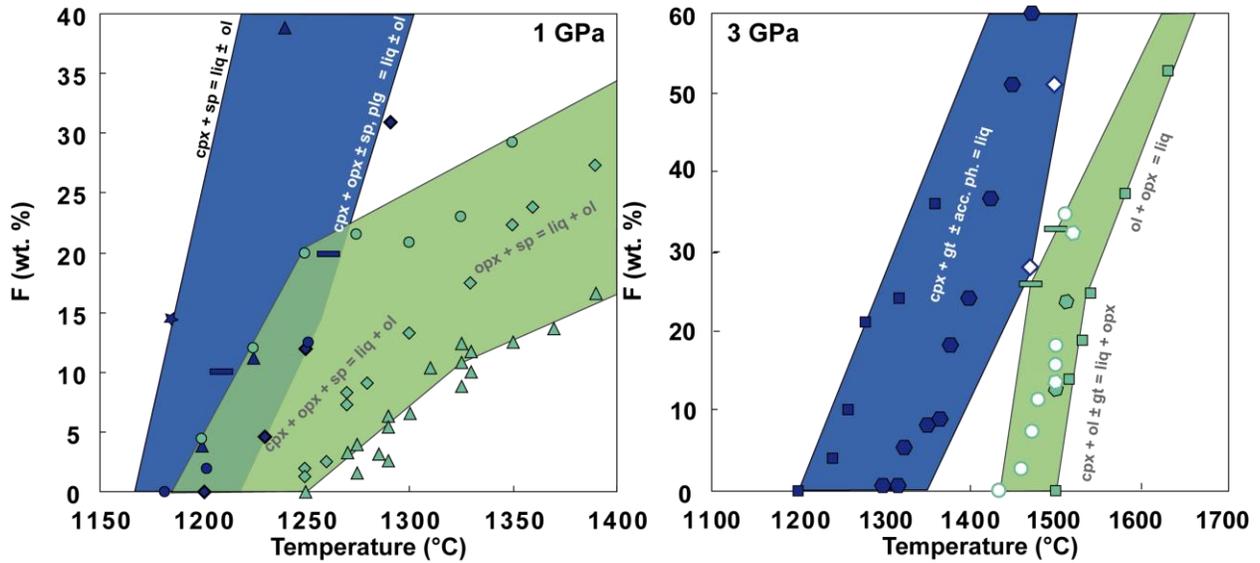

**Figure 2.** Experimental melt fractions as functions of temperature for pyroxenites (blue: M7-16, star; M5-40, triangles; M5-103, diamonds; GV10, rectangles; Ito and Kennedy's eclogite, circles; GA2, squares; G2, hexagons; B-ECL1, open diamonds) ) and peridotites (green: PHN-1611, circles; MM3, diamonds; DDMM, triangles; KR4003, square; KLB-1, open circles; HK66, rectangles; KG2, hexagons; KG1,open hexagons) at 1 and 3 GPa. The blue and green fields show the envelopes of melt fractions for pyroxenites and peridotites, respectively. References: Baker and Stolper (1994), Baker et al., (1995), Borghini et al. (2017), Hirose and Kushiro (1993), Hirschmann et al. (1998), Ito and Kennedy (1974), Kogiso and Hirschmann (2006), Kogiso et al. (1998), Kushiro (1996), Lambart et al. (2009), Pertermann and Hirschmann (2003a), Spandler et al. (2008), Walter (1998), Wasylenki et al. (2003).

Peridotites are marked by a decrease in melt productivity when cpx disappears from the experimental assemblage. However, the productivities of pyroxenites and peridotites are scattered but largely overlap before cpx disappearance. Hence, the cpx disappearance in the mineralogical assemblage is the main parameter influencing the large difference in melt productivity between peridotites and pyroxenites.

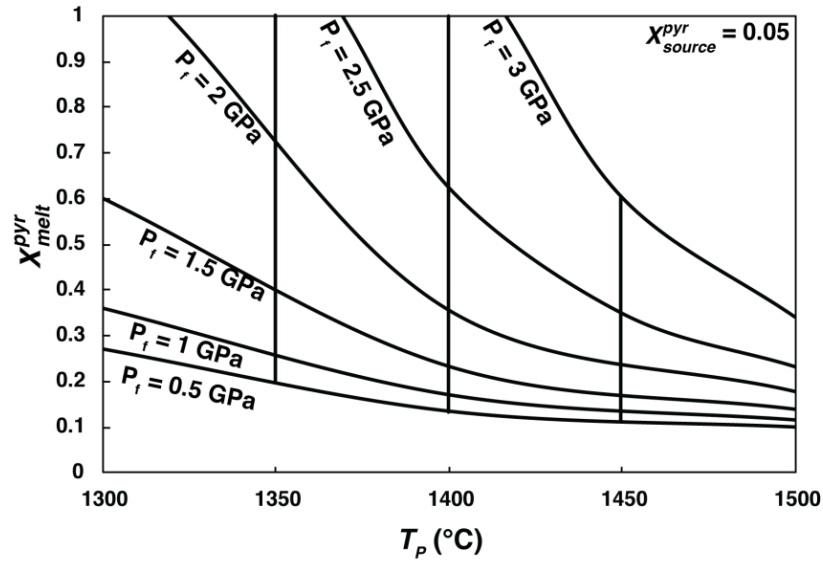

**Figure 3.** Effect of potential temperature ($T_P$) and the final pressure of melting ($P_f$) on the contribution of pyroxenite G2 (Pertermann and Hirschmann, 2003a) in the aggregated magma $X_{melt}^{pyr} = (X_{source}^{pyr} \times F_f^{pyr})/(X_{source}^{pyr} \times F_f^{pyr} + (1 - X_{source}^{pyr}) \times F_f^{per})$, assuming $X_{source}^{pyr} = 0.05$ and calculated with Melt-PX (Lambart et al., 2016).

Finally, the thermal structure of the mantle and the crust mostly controls the potential temperature ($T_P$) and the thickness of the lithosphere (considered as the final pressure of melting; e.g., Dasgupta et al., 2010). Decreasing $T_P$ or increasing the thickness of the lithosphere both result in an increase of the low-solidus component contribution (usually, the pyroxenite) in the melt by decreasing the relative height of the melting column where both lithologies are melting **(Fig. 3)** (Ellam, 1992; Humphreys and Niu, 2009; Niu et al., 2011; Sobolev et al., 2007)**.**

## 2.2 Peridotite vs. pyroxenite: experimental melt compositions

### 2.2.1. Major elements

Pyroxenites are usually separated into two groups: silica-excess (SE) and silica-deficient (SD) pyroxenites. The garnet–pyroxene thermal divide separating these two groups (defined by the Enstatite - Diopside - Alumina plane in the $CaO-MgO-Al_2O_3-SiO_2$ tetrahedron; O'Hara, 1976) controls the melting relations of pyroxenites at high pressure ($\geq$ 2 GPa; Kogiso et al., 2004) when cpx and garnet are both present in the mineralogical assemblage. Lambart et al. (2013) reviewed the melting phase relations of pyroxenites and showed that there is a large overlap of compositions between melt produced by the different mantle lithologies. As a rule, there is a progressive transition from the liquids derived from SE compositions to SD compositions to peridotites rather than a sharp compositional contrast. The extreme variability of the pyroxenite melts and the progressive transition between melts from different lithologies make finding good markers for the presence of pyroxenites in the source challenging. Yang and Zhou (2013) and Yang et al. (2016) suggested that the FC3MS ($FeO/CaO-3*MgO/SiO_2$, in wt.%) parameter can help with distinguishing between pyroxenite- and peridotite-derived melts. Their analysis demonstrates that the upper boundary of the FC3MS value for peridotite melts is 0.65, while melts derived from pyroxenites cover a much larger range (-0.9 to 1.7).

More recently, Yang et al. (2019) proposed two new parameters (FCMS and FCKANTMS) and showed that using log ratio-based chemical markers help in reducing the temperature and pressure effects on the compositional heterogeneity of melts. However, we note that none of these parameters can help to distinguish between peridotite-derived and pyroxenite-derived melts for compositions producing low parameter values (i.e, mostly SD pyroxenites).

### 2.2.2. First Row Transition elements.

First Row Transition Element (Sc, Ti, V, Cr, Mn, Fe, Co, Ni, Cu, Zn) concentrations in magmas have been recently suggested as good indicators of the mantle source mineralogy (e.g., Davis et al., 2013; Herzberg, 2011; Humayun et al., 2004; Le Roux et al., 2010, 2011; Prytulak and Elliott, 2007; Qin and Humayun, 2008; Sobolev et al., 2005, 2007). Le Roux et al. (2011) showed that during partial melting of the upper mantle, olivine and orthopyroxene do not significantly fractionate Mn and Fe from each other and melts from garnet-free peridotite are expected to have similar Mn/Fe ratios as the source. In contrast, clinopyroxene and garnet, the dominant minerals in pyroxenitic assemblages (Kogiso et al., 2004), show strong fractionations, such that melts of pyroxenites or eclogites would be expected to produce melts with low Mn/Fe compared to peridotite partial melts (Humayun et al., 2004). However, experiments by Le Roux and coworkers only explored peridotitic mineral compositions at a limited range of pressure (1.5-2 GPa), while partition coefficients can vary depending on the compositions of the minerals and melts (e.g., Davis et al., 2013; Wang and Gaetani, 2008) as well as with pressure and temperature (P-T) conditions (e.g., Matzen et al., 2013, 2017a).

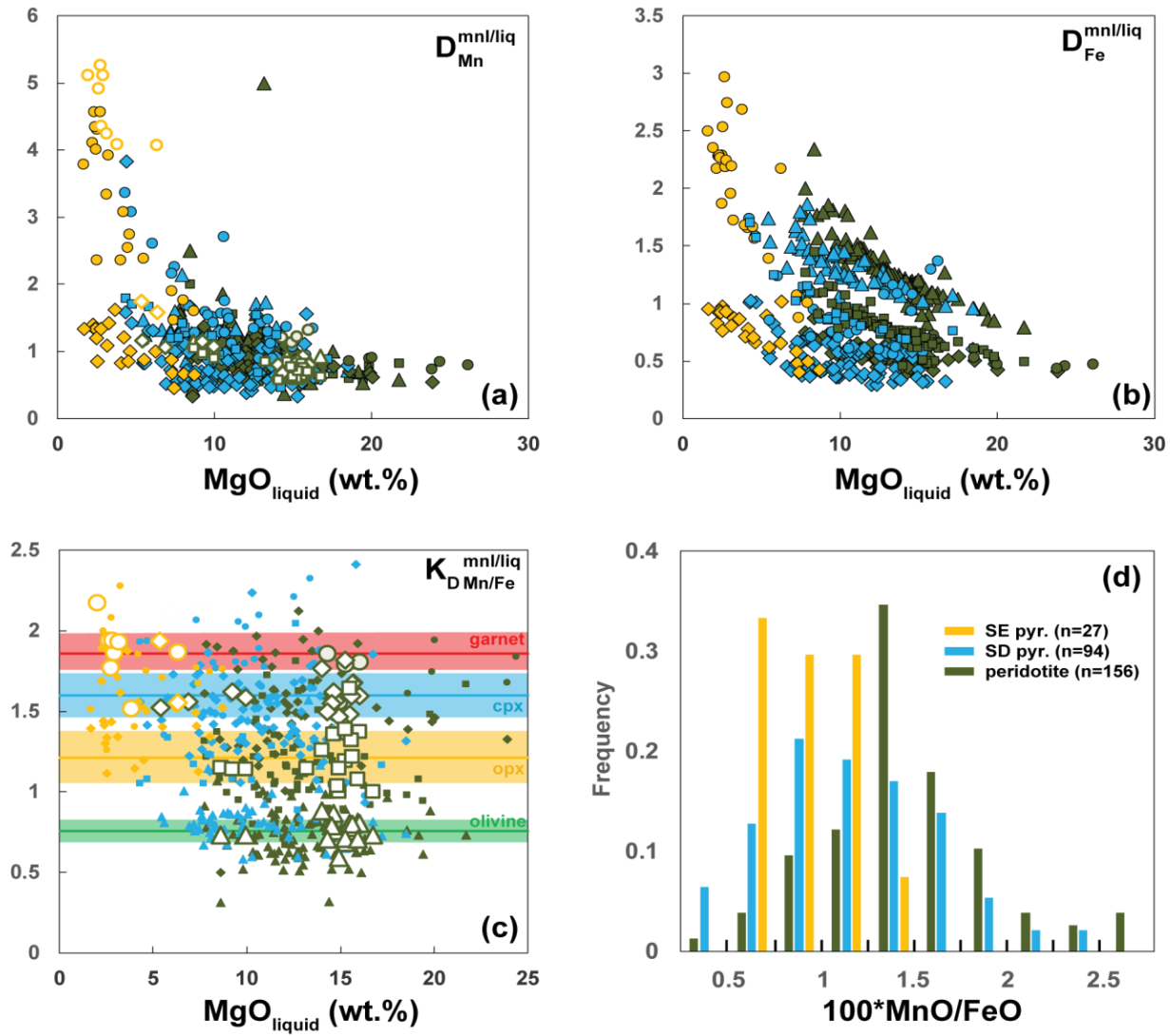

**Figure 4.** (a-b) Partition coefficient of Mn (a) and Fe (b) between minerals (garnet: circles; cpx: diamonds; opx: squares; olivine: triangles) as functions of the MgO content of the liquid for pyroxenites (SE: yellow; SD: blue) and peridotites (green). In (a), open symbols are higher accuracy analyses obtained by LA-ICP-MS or high current microprobe analyses (Davis et al., 2013; Le Roux et al., 2011; Pertermann et al., 2004). (c) Mn-Fe exchange partition coefficients ($K_D$) between minerals and melts. The colored lines show the average $K_D$ calculated for each mineral using high accuracy measurements and the colored bands show the corresponding standard deviation. (d) Histogram showing the distributions of the 100*MnO/FeO ratios in experimental melts. References: Baker and Stolper (1994), Borghini et al. (2017), Davis et al. (2013), Falloon and Danyushevsky, (2000), Keshav et al., (2004), Kogiso and Hirschmann (2001, 2006), Kogiso et al. (2003), Lambart et al., (2009; compositions of the solid phases can be found at https://doi.org/10.6084/m9.figshare.9926885.v1), Lambart et al. (2013), Laporte et al. (2004), Le Roux et al. (2011), Pertermann and Hirschmann (2003b), Pickering-Witter and Johnston (2000), Pilet et al., (2008), Schwab and Johnston (2001), Walter (1998), Wasylenki et al. (2003), Yasuda et al. (1994), Yaxley and Sobolev, (2007).

To take into account the potential effects of compositional and P-T variations, we plotted the partition coefficients between melt and mineral compositions using experiments reported in the literature on SE and SD pyroxenites and on peridotites. Partition coefficients for Fe and Mn strongly depend on the MgO content of the melt **(Fig. 4a,b)**, but the Mn–Fe $K_D$ are distinct for each mineral, and cpx-melt and garnet-melt Mn-Fe $K_{DS}$ are distinctly higher than olivine-melt and opx-melt $K_{DS}$ **(Fig. 4c)**. In addition to this, and in agreement with the results from Le Roux et al. (2011), Mn-Fe $K_{DS}$ are independent of the composition. These results support the conclusions of Le Roux and coworkers: a mineral assemblage dominated by cpx and garnet will produce a melt with lower Mn/Fe ratio than a mineral assemblage dominated by olivine and orthopyroxene and, despite the large variability of bulk compositions that obscures this relationship, Mn/Fe ratio produced by pyroxenite-derived melts are generally lower than that in peridotite-derived melts **(Fig. 4d).**

## 3 Interactions between chemical heterogeneities and peridotite

Solidi of the nominally anhydrous subducted oceanic crust (e.g., Pertermann and Hirschmann, 2003a) and sediments (e.g., Spandler et al., 2010) are at lower temperatures than the solidus of nominally anhydrous peridotite-bearing ambient mantle **(Fig. 5a).** This implies that when the ambient peridotitic mantle carrying chemical heterogeneities upwells along an adiabat in the upper mantle **(Fig. 5b),** the heterogeneities reach their solidi temperatures deeper than the peridotite. Thus, between the solidi of heterogeneities and the peridotite (indicated by the dashed areas in **Figs. 5a** and **b**), the partial melt from heterogeneities would encounter subsolidus (unmelted) peridotite. The presence of volatiles such as $H_2O$ and/or $CO_2$ in the heterogeneities further depress the solidi to lower temperatures (e.g. Dasgupta et al., 2004; Hammouda, 2003; Hermann and Spandler, 2008; Poli and Schmidt, 1995; Skora and Blundy, 2010; Yaxley and Brey, 2004) implying that the heterogeneities begin to partially melt even deeper in the upper mantle while the surrounding peridotite is still subsolidus. (An exception to this is under highly oxidized regions where $CO_2$ in peridotites may cause onset of carbonated silicate partial melting deeper than in the case of volatile-free peridotite (Dasgupta et al., 2013)).

The partial melts from these heterogeneities are not in equilibrium with subsolidus peridotites and will react with them (e.g., Kogiso et al., 1998; Mallik and Dasgupta, 2012; Yaxley and Green, 1998). Such a reactive melt-rock interaction can be illustrated by a simple binary phase diagram between olivine and quartz with double-eutectic at relevant pressures **(Fig. 5c).** (The readers are reminded that the binary phase diagram, while a useful tool to illustrate the first-order reactive crystallization process, does not capture the complexities imposed on the phase equilibria by a multi-component system that is a better analog to natural systems.) The siliceous partial melts from the heterogeneities lie on the liquidus surface of the pyroxene-quartz binary, while the peridotite lies below the solidus of the olivine-pyroxene binary. When the siliceous partial melts react with peridotite, the bulk composition of the melt-peridotite mixture is located between the partial melt and peridotite. Whether the bulk composition lies in the pyroxene-quartz binary or the olivine-pyroxene binary depends on the melt-rock proportion, where a higher proportion would place the bulk composition closer to the pyroxene-quartz binary. Thus, the interaction between the partial melt of the heterogeneities and peridotite would result in reactive crystallization, as observed in the phase diagram. The first order reaction that takes place in this context is as follows:

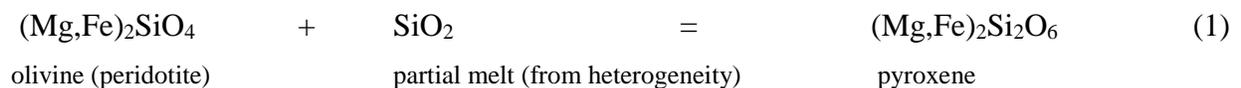

$$(Mg,Fe)_2SiO_4 \quad + \quad SiO_2 \quad = \quad (Mg,Fe)_2Si_2O_6 \quad (1)$$

olivine (peridotite)       partial melt (from heterogeneity)         pyroxene

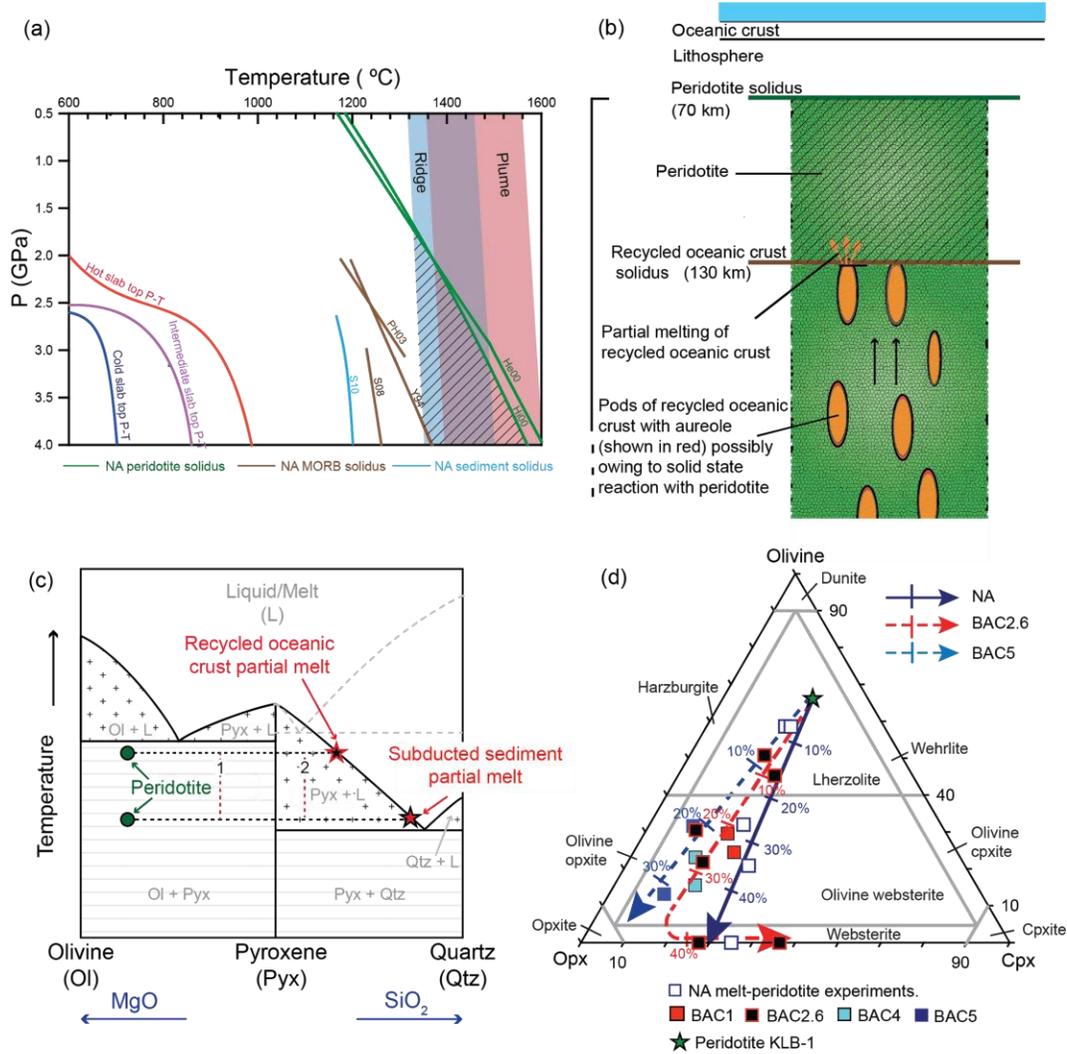

*Figure 5. (a)* Solidus of nominally anhydrous (NA) peridotite, recycled oceanic crust (MORB) and sediment plotted in pressure-temperature space. Also plotted for comparison are the range of adiabats from mid-ocean ridges to plume-settings with the lower end bracketed by mantle potential temperature of 1315 °C (McKenzie et al., 2005) and the higher end by 1550 °C (Herzberg et al., 2007), as well as hot, intermediate and cold subducted slab geotherms (Syracuse et al., 2010). The dashed area indicates P-T space where partial melt from recycled oceanic crust and sediments would react with the surrounding subsolidus peridotite. *(b)* Cartoon showing a slice of upwelling peridotitic mantle with pods of lithological heterogeneity (e.g. 5 wt.% recycled oceanic crust). NA recycled oceanic crust encounters its solidus at a depth ≈ 130 km, while the NA peridotite is still below the solidus. In the dashed area (the same as in panel a), partial melt derived from recycled oceanic crust undergoes reactive infiltration through the surrounding subsolidus peridotite. *(c)* Binary phase diagram between olivine and quartz (not to scale) with double eutectic, at 2.5 GPa (Chen and Presnall, 1975). Dashed gray lines in the pyroxene-quartz binary represent the solidus and liquidii in a pure $MgO$-$SiO_2$ system, solid lines represent solidus and liquidii with alkalis included. '1' indicates a bulk composition with low melt-rock ratio resulting in complete consumption of the melt and producing olivine-pyroxene residue. '2' indicates a bulk composition with high melt-rock ratio which produces a pyroxene residue by consuming olivine, and a residual melt that is less siliceous and more magnesian than the melt directly derived from chemical heterogeneities. *(d)* Streikeisen diagram for ultramafic rocks. The peridotite KLB-1 (Takahashi, 1986) is representative of a shallow upper mantle peridotite. The other symbols plotted are residues of reactive crystallization between peridotite and melt derived from low to moderate degree of partial melting of recycled oceanic crust with various proportion of $CO_2$, from 0 wt.% (NA, Mallik and Dasgupta, 2012) to 5 wt.% (BAC, Mallik and Dasgupta, 2014, 2013) at 2.5 - 3 GPa. The percentage next to the symbols represent the initial proportion of melt to peridotite.

A high melt to rock proportion would result in the consumption of olivine and melt to crystallize pyroxene, with remaining partial melt that is less siliceous and more Mg-rich than the partial melt of the heterogeneity. A low melt to rock proportion would result in complete crystallization of the partial melt to form pyroxene, with residual olivine from the peridotite. Similar reasoning can be made with other heterogeneity compositions: silica deficient pyroxenites will result in olivine production at the expense of orthopyroxene, but for a low melt-rock ratio, the melt will be consumed during the reaction (Lambart et al., 2012). Thus, such a process of reactive crystallization between partial melt from a heterogeneity and peridotite not only produces distinct residual partial melt compositions, but also contributes to further heterogeneity in the upper mantle **(Fig. 5d)**. Due to the interaction of heterogeneities with peridotite, such melt-rock reactions may occur beneath mid-ocean ridges (Borghini et al., 2017; Lambart et al., 2012), ocean islands (Mallik and Dasgupta, 2012, 2013, 2014; Sobolev et al., 2005), continental lithosphere (Chen et al., 2017; Kiseeva et al., 2013), in subduction zones near depths of arc magma generation (Kelemen, 1995; Mallik et al., 2015, 2016; Prouteau et al., 2001; Tatsumi, 2001; Wang and Foley, 2018), and as deep as the transition zone (Kiseeva et al., 2013; Thomson et al., 2016).

**4 Sampling melts of the mantle - approach and overview**

In this section, we present a brief, and by no means comprehensive, survey of the magma generation processes and compositions of natural samples from three tectonic settings on Earth: mid-ocean ridges, ocean islands, and volcanic arcs. For each setting, we review experimental constraints on magma genesis, followed by a discussion of the compilation of natural compositional data. To facilitate as many comparisons between experimental data and natural data across the three settings, we focus exclusively on major element compositions. Obviously, using only major elements rather than emphasizing only primitive samples presents limitations for fingerprinting the nature of the melting source. Our goal here is not to identify the source characteristics, since trace element ratios and isotopes would be better suited for that task, but to make broad comparisons between experimental, modeling, and natural observations in terms of petrogenetic processes across the major tectonic settings of magmatism on Earth.

**Supplementary Data** and literature sources used in this review are summarized in Table S1 and are all reported in **Supplementary Tables S6, S7 and to S8**. Owing to the extremely large number of samples, particularly for arc volcanic rocks (nearly 40,000), we contoured all the volcanic rock data using a smoothed histogram method (Eilers and Goeman, 2004), using 75 histogram bins in the x and y direction and a positive smoothing parameter (λ) of 1.5 (higher values of λ lead to more smoothing; a value near zero would result in a plot of essentially the raw data). Density contours for volcanic rocks in **Fig. 6** were constructed at intervals of 10%. We applied the same parameters in the contouring for the volcanic rock data from each tectonic setting so side-by-side comparisons can be accurately made. Owing to the small number of plutonic cumulate rocks (<500 at any given setting) and melt inclusions (<200, from volcanic arcs only), we did not contour these datasets.

4.1 MORB

*4.1.1 Intracrustal processing.*

The relatively simple geodynamic setting of mid-ocean ridges serves as a baseline model for mantle melting and magma generation. A primary magma is one formed during adiabatic decompression melting of mantle peridotite and which has not been modified during ascent and eruption. It has long been recognized that such a magma is rarely, if ever, sampled at mid-ocean ridges (or any tectonic setting for that matter) – the average Mg# of all MORB is 56, significantly lower than the value of ~70 expected for equilibrium with mantle peridotite. Three main factors are involved in producing the compositional spectrum of mid-ocean ridge basalts (MORBs): 1) mantle source heterogeneity, 2) style of mantle melting, and 3) intracrustal processing. In this section, we focus on the role of intracrustal processing, which marks the boundary between the melting regime and the onset of crystallization, differentiation, eruption and eventual sampling.

In order to go from sampled MORBs from the seafloor to how such MORBs looked when they parted ways with the mantle, the contributions from crystallization-differentiation, which occur in the crust must first be understood. Intracrustal processing of MORB can be broadly considered as variations on the theme of fractional crystallization (FX). The simplest scenario, and, today, the paradigm, is that MORB follow a differentiation path under anhydrous, low pressure conditions, evolving towards the classic Fe-enrichment (tholeiitic) trend (O'Hara, 1965). Experiments and observations of natural MORBs indicate the generalized low-pressure crystallization sequence of olivine → olivine + plagioclase → olivine + plagioclase + clinopyroxene → plagioclase + clinopyroxene + orthopyroxene + Fe-Ti oxides (Coogan, 2014; Grove et al., 1993). However, detailed study of MORB over the past decades reveal several inconsistencies with the simple FX model: 1) phenocryst assemblages in MORB are not representative of cotectic proportions as predicted by FX (Francis, 1986), 2) FX models require fractionation of clinopyroxene, but it is not commonly observed as a phenocryst, and 3) some incompatible elements, like Th, are overly enriched in global MORB compared to predictions from simple fractional crystallization (Coogan and O'Hara, 2015; Jenner and O'Neill, 2012). Thus, more complex scenarios have been hypothesized to explain the mismatch between trace element data and simple fractional crystallization models. The two major variations on the simple fractional crystallization model are 1) in situ crystallization, where crystallization of a magma occurs in a progressive solidification front and involves return of interstitial melt to the magma chamber (Coogan and O'Hara, 2015; Langmuir, 1989; McBirney and Noyes, 1979) and 2) "RepTapFrac", wherein a magma chamber is periodically replenished (Rep), melt extracted (Tap), and then fractionally crystallized (Frac) (Albarede, 1985; O'Hara, 1977; O'Neill and Jenner, 2012). Additionally, recent studies suggest an alternative – reactive porous flow/melt-rock reaction – as another mechanism that could reconcile several of the MORB conundrums, particularly the observed over-enrichment in incompatible trace elements (Coumans et al., 2016; Gao et al., 2007; Lissenberg and MacLeod, 2016; Lissenberg et al., 2013). Although trapped melt and melt-rock reaction (Lissenberg and Dick, 2008) in the lower crust and even the oceanic mantle lithosphere are increasingly recognized as additional processes contributing to MORB compositional evolution (especially in the case of incompatible trace elements), the evolved nature of MORB (Mg# ~56) compared to primary mantle melts (Mg# ~70) and the overall fractionation of incompatible elements between lower oceanic crust and upper oceanic crust (Coogan, 2014) still broadly support a fundamental role for crystal-liquid separation in MORB petrogenesis.

*4.1.2 Natural MORBs & their cumulates.*

The perspective on MORB and oceanic crust generation has been largely focused on extrusive rocks due to accessibility – there are several thousands of samples of MORB (pillow lavas, glasses) but far fewer samples of complementary subcrustal MORB cumulates **(Table S5).** A coupled perspective of both MORBs and their cumulates can also be illustrative of the factors involved in MORB evolution. In **Figure 6a-d**, we plot both global MORBs (both lavas and glasses from the database of (Gale et al., 2013)) and a recently compiled global MORB cumulate database (Chin et al., 2018) in Harker-style diagrams. Compared to other tectonic settings (discussed in later sections), the combined global database of MORBs and MORB cumulates collectively fall on tight, minimally scattered compositional trends. Throughout a wide range of Mg# (and thus differentiation index), the $SiO_2$ content of MORBs remains relatively constant at ~55 wt.% **(Fig. 6a).** Crystal fractionation of olivine, followed by olivine + albitic plagioclase, is apparent in plots of $Al_2O_3$ vs. Mg# **(Fig. 6b)** and $Na_2O$ vs. Mg# **(Fig. 6c).** The consistently low $K_2O$ vs. Mg# trend of MORB cumulates **(Fig. S1)** also suggests that crystal-liquid segregation is generally efficient (Natland and Dick, 1996). Finally, the classic tholeiitic magma series is revealed in the cumulate counterparts of MORBs – the Fe-enrichment trend of MORB is balanced by the Fe-poor nature of early MORB cumulates, and the late precipitation of Fe-Ti oxides in oxide gabbros is clearly shown in a marked increase in $TiO_2$ at Mg# ~50 **(Fig. 6d)** (Chin et al. 2018). To first order, major element systematics of MORB + MORB cumulates supports crystal-liquid segregation along tholeiitic differentiation trends. Yet, additional observations, such as reaction textures and complex zoning patterns (Lissenberg and MacLeod, 2016), isotopic heterogeneity (Lambart et al., 2019), and phase assemblage disequilibrium (Gillis et al., 2014) in associated cumulates, seem to prevent petrologists from establishing a clear genetic link between them and MORB exclusively through fractional crystallization. While a summary of trace element systematics of MORB and MORB cumulates is beyond the scope of this study, conflicts persist as to the roles of melt-rock reaction, secondary melt infiltration - as shown by trace element studies - versus fractional crystallization, which is broadly supported by major element trends.

*4.1.3 Primary MORB.*

Co-variations observed in MORB suites of isotopic ratios of heavy elements (e.g., Sr, Nd, Pb, Os, Hf), being insensitive to melting and fractionation processes, are usually attributed to mantle heterogeneity (Hofmann, 2003). Additionally, timing constraints from U-series disequilibria suggest that melt extraction from the mantle is rapid (Elliott & Spiegelman, 2003; Elkins et al., 2019 and references therein). It has been suggested that high-permeability channels from the melt source region toward the surface must exist (Kelemen et al., 1995) to satisfy these constraints. Numerical models (e.g., Weatherley and Katz, 2012, 2016) show that the formation of these channels would be facilitated by the presence of lithological heterogeneities in the mantle owing to thermal interaction between partially molten and subsolidus lithologies.

To decipher source signals, MORB need to be corrected for crystal fractionation. Once corrected for low-pressure fractionation (Klein and Langmuir, 1987), MORB show evidence for variability in major-element compositions, likely due to compositional (e.g., Niu and O'Hara, 2008) or thermal (e.g., Dalton et al., 2014; Langmuir et al., 1992; Gale et al., 2014) variability in the source. However, Neave et al. (2019) recently showed that the major element composition of a primitive magma also affects the efficiency with which it crystallises.

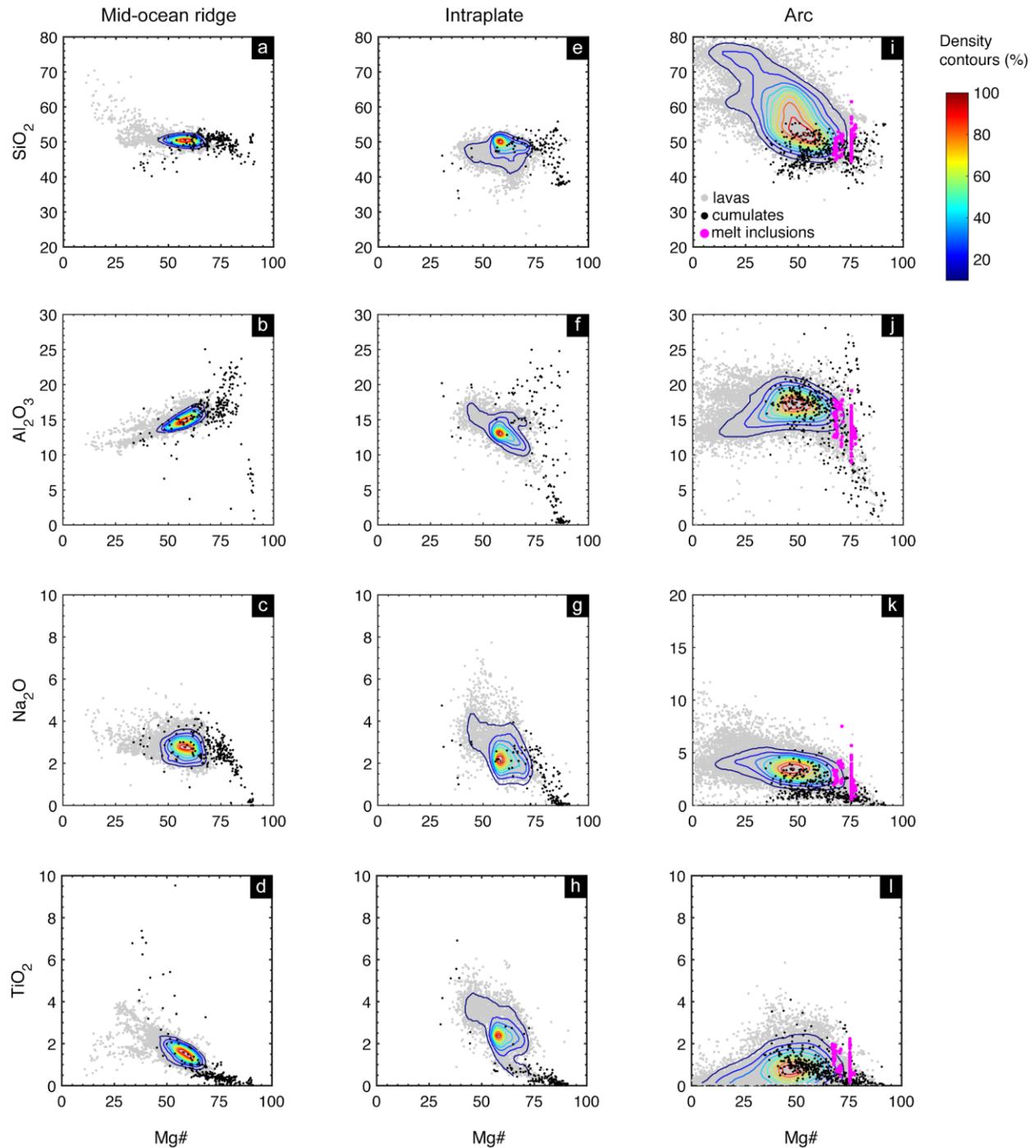

**Figure 6.** Major element oxides (wt.%) vs. Mg# for lavas (gray circles), cumulates (black circles), and melt inclusions (magenta symbols, from volcanic arcs only). Colored curves show the density contours (at intervals of 10%) for volcanic rocks.

Additionally, Bennett et al. (2019) demonstrated that the vertical extent of mid-ocean-ridge magmatic systems is not restricted to the first few kilometers as previously thought, but could extend down to the base of the lithospheric mantle. Thus, predominantly shallow-level fractional crystallization of MORB may be more complex than commonly thought. Finally, in comparison to ocean island and arc basalts, the range of primary MORB major-element compositions stays restricted. This restricted variability can be explained by significant low-pressure magma mixing (e.g., Shorttle, 2015; Lambart et al., 2019), and/or by similar melting reactions and melt compositions generated by pyroxenites and peridotites in the melting zone beneath mid-ocean ridges (e.g., Lambart et al., 2009).

It is beyond the scope of this short section to discuss all of these contributing factors to MORB geochemistry in detail and we refer the reader to comprehensive reviews on MORB petrogenesis in the literature (Gale et al., 2014; Langmuir et al., 1992; Lissenberg et al., 2019; Niu and O'Hara, 2008; Rampone and Hofmann, 2012; White and Klein, 2014).

4.2. Oceanic Island Basalts

Ocean island basalts (OIB) have received considerable attention in the literature given their potential as a proxy to unravel the composition and thermal state of the mantle. Isotopic and trace element diversity in ocean island basalts are indicative of chemically heterogeneous sources such as the presence of enriched and depleted mantle reservoirs, recycled oceanic crust, continental crust and sediments. The readers are referred to the seminal reviews by Hofmann (1997, 2003) for details on how geochemical characteristics of oceanic basalts give us insight into the dynamics of the mantle.

*4.2.1 Source lithologies of primitive OIB*

Evidence for presence of recycled material in the mantle source of OIB come from major-, trace-element and isotopic compositions. In fact, previous studies have argued that the high $TiO_2$ contents (> 2.5 wt.%) of primary OIB are indicative of the presence of recycled oceanic crust in the source (**Fig. 6f**; Prytulak and Elliott, 2007). In addition, OIB compositions extend to very low silica contents (**Fig. 6e**) and primary magma with very low $SiO_2$ concentrations (≤ 46 wt.%) also display the highest $^{206}Pb/^{204}Pb$ ratios, indicative of recycled crust in the source (Jackson and Dasgupta, 2008). Such low $SiO_2$ contents are observed in experimental partial melts of recycled heterogeneities such as silica-deficient pyroxenites (Hirschmann et al., 2003; Kogiso et al., 2003; Lambart et al., 2009) and carbonated oceanic crust (Mallik and Dasgupta, 2014). However, most partial melts of recycled heterogeneities do not reach Mg# 73 (Kogiso et al., 2004; Pertermann and Hirschmann, 2003b). Such high Mg# can be explained if both pyroxenite- and peridotite-derived melts contribute to OIB genesis or if the pyroxenite-derived melt reacts with the peridotite before directly contributing to magma genesis (Sobolev et al., 2005,2007; Mallik and Dasgupta, 2012, 2013, 2014; Phipps Morgan, 2001). In fact, the process of reactive crystallization as a consequence of such a reaction between pyroxenite (eclogite)-derived melt and peridotite is described in Section 3 and results in MgO enrichment and $SiO_2$ depletion of the eclogite melt. In summary, it is widely accepted that the mantle source of OIB is heterogeneous, but the lithological nature and chemical composition of this heterogeneity is still debated (e.g., eclogite, pyroxenite, hybrid lithologies, refertilized peridotites). However, the compositions of the various mantle components control their contribution to magma genesis by affecting their melt productivity (Lambart et al. 2016) and reactivity with the surrounding mantle (Lambart et al., 2012; Mallik and Dasgupta, 2012, 2014; Sobolev et al., 2005).

To illustrate the importance of the mode of participation of the heterogeneities in the source of OIBs and their compositions, we perform an exercise as described below. We remind the readers that the primary goal of this exercise is to highlight the effects of the choice of the melting model, and less to quantify the proportion of pyroxenite in the mantle source of OIB. We first consider a simple case where deeper melt derived from recycled crust is channelized through the peridotite matrix (dashed region in **Fig. 5a, b**) and mixes with peridotite partial melt at shallower depths to produce the observed primary magma compositions. We call this the 'melt-melt mixing model'. In this model, melt-rock interactions are not taken into account. We consider two pyroxenite compositions: G2, a MORB-type eclogite (Pertermann and Hirschmann, 2003a) and MIX1G, the average natural pyroxenite (Hirschmann et al., 2003), and estimate their proportions in the sources of ocean islands globally as follows: we use the primary melt compositions and LAB (Lithosphere – Asthenosphere boundary; proxy for the final depth of melting) thicknesses at the time of volcanism for each individual ocean island as reported by Dasgupta et al., (2010) and we use PRIMELT3 MEGA.XLSM software (Herzberg and Asimow, 2015) to estimate the potential temperature ($T_P$) for each island. As described in Section 2, the melts derived from eclogitized crust (or any lithologies dominated by clinopyroxene and garnet) would have a lower Mn/Fe than partial melts of peridotite or high Mg-pyroxenite containing olivine and/or orthopyroxene. Thus, olivines that would crystallize from a melt with contributions from clinopyroxene and garnet bearing pyroxenites would also mirror the low Mn/Fe signatures. Sobolev et al., (2007) used this observation to derive an empirical relationship between Mn/Fe of olivines in equilibrium with primary basalts and the contribution of pyroxenite in the aggregate basaltic melts ($X_{melt}^{pyr}$):

$$X_{melt}^{pyr} = 3.48 - 2.071 \times (100 Mn/Fe) \quad (2)$$

We note that Matzen et al. (2017b) demonstrated that the Mn content variation in early crystallizing olivines can be explained by partial melting of peridotite over a range of pressures, and that recycled lithologies may be present in the source but are not required to explain the Mn variations in olivines. However, the goal of the following exercise is to simply demonstrate that the estimated proportion of pyroxenite in the mantle source of OIBs varies with the choice of melting model and composition of pyroxenite.

With the PRIMELT3 MEGA.XLSM software, we calculated the Mn/Fe ratio of the olivine in equilibrium with the primary OIB and used Sobolev's proxy to estimate $X_{melt}^{pyr}$ from the ocean islands. Finally, the proportions of pyroxenites in the source of the ocean islands ($X_{source}^{pyr}$) are calculated using Melt-PX (Lambart et al., 2016) and assuming the following relation that includes a first-order assumption that the proportion of pyroxenite in the melt is weighted by the abundance of pyroxenite in the source:

$$X_{melt}^{pyr} = \left(X_{source}^{pyr} \times F_f^{pyr}\right) / \left(X_{source}^{pyr} \times F_f^{pyr} + \left(1 - X_{source}^{pyr}\right) \times F_f^{per}\right) \quad (3),$$

with $F_f^{pyr}$ and $F_f^{per}$, the final degree of melting of the pyroxenite and the peridotite, respectively, when the parcel of upwelling mantle reaches the LAB.

In a second set of calculations, we use the same constraints on $T_P$ and $P_f$, but consider that melt derived from G2 first reacts with the peridotite before contributing to magma genesis (e.g., Sobolev et al., 2005). We call this the 'reactive crystallization model'. Mallik and Dasgupta (2014) used published experimental results to derive an empirical relation that predicts the major element concentrations in a primary melt that forms by reactive crystallization of partial melts from eclogite-derived crust (similar to G2) as they undergo porous infiltration through a subsolidus peridotite matrix. While the proxy of Sobolev et al. (2007) is based on the trace element compositions in olivine (equation 1), Mallik and Dasgupta's model is based on the major element composition of the primary magmas. Using two independent models allows us to test for the consistency of such empirical relationships. In Mallik and Dasgupta's model, melt-derived from the recycled crust reacts with the peridotite to form a hybrid lithology and a distinct melt. The distinct melt is in chemical and thermal equilibrium with the hybrid lithology, and this melt, by mixing with peridotite-derived melt results in the observed primary magma composition. The empirical relation of Mallik and Dasgupta provides the fraction of recycled crust-derived melt required to produce the new hybrid lithology ($X^{rl}$), and the proportion of melt derived from the hybrid lithology ($X_{melt}^{hyb}$). Using these, we can estimate the proportion of G2 (before melting) in the source using the following equation (Sobolev et al., 2005)

$$X_{source}^{G2} = \frac{X^{rl}}{F^{G2} \times \left( \frac{1-X_{melt}^{hyb}}{X_{melt}^{hyb}} \times \frac{F_f^{hyb}}{F_f^{per}} + \frac{1-F^{G2}}{F^{G2}} \times X^{rl} + 1 \right)} \qquad (4),$$

where, $F^{G2}$ (= 8.9 wt.%) is the melting degree undergone by G2 before reacting with the surrounding mantle, and $F_f^{hyb}$ and $F_f^{per}$, the melting degrees of the hybrid lithology and the peridotite calculated with Melt-PX when the mantle parcel reaches $P_f$. **Fig. 7** presents estimated $X_{source}^{pyr}$ for each major group islands in the three oceans for the two models of melting. Calculations have been performed on individual islands listed in Dasgupta et al. (2010) (see **Supplementary Table S9**). To facilitate the comparison, we report the results averaged by island groups and the error bars show the variability between islands from the same group. Hence, a group with only a single island, such as Iceland, has no error bars. The proportion of pyroxenites in the mantle source significantly varies between the type of calculations, but also between and within each island group. Using the 'melt-melt mixing model' and the composition G2, the average proportion of pyroxenite in the source contributing to OIB genesis is ~34 % and varies from 8% (Balleny) to 100% (Tristan da Cunha and Gough). Due to its higher solidus temperature than G2 (Lambart et al., 2016), using the composition MIX1G results in an overall higher proportion of pyroxenite in the source required to explain the Mn/Fe ratio (~54% in average). Proportions of G2 estimated with the 'reactive crystallization model' vary from 0% (Iceland and Amsterdam) to 61% (Cameroon and Fernando de Norhona). In summary, despite constraints on the mantle temperature and final pressure of melting, the choice of the model and composition of the enriched component/heterogeneity will significantly affect the estimates of its proportion in the mantle source of magmas. Despite these discrepancies, these models converge to the conclusion that a large fraction [39 (± 28)%] of recycled material is present in the mantle source of OIB .

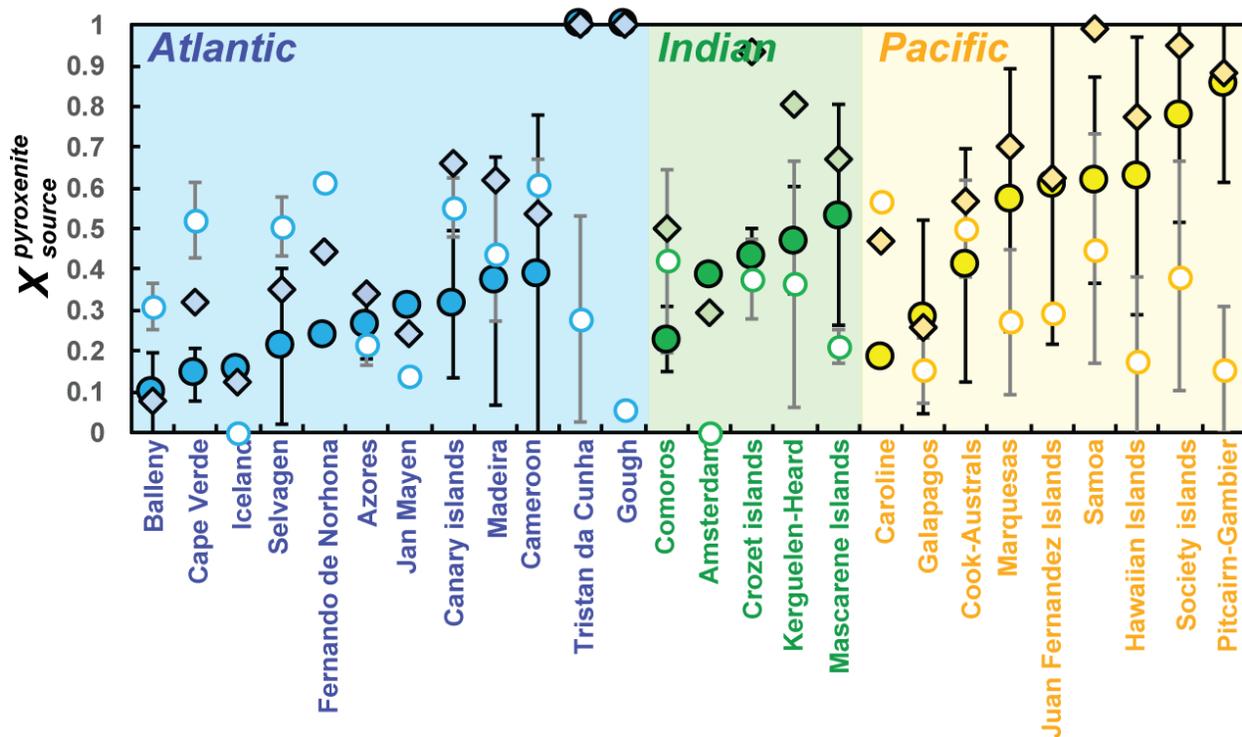

**Figure 7.** Calculated average proportions of pyroxenite in the source of the main island groups using a simple melting model ('melt-melt mixing model') with either G2 (filled circles) or MIX1G (diamonds) as pyroxenitic component and using a 'reactive crystallization model' using G2 (open circles) as pyroxenitic component. The error bars on G2 fractions correspond to one standard deviation on each island group calculated from fractions obtained for individual island (Table S9). (Black and grey error bars show variability inside each island group calculated using the melt-melt mixing model with G2 and using the reactive crystallization model, respectively).

We also note there is no strong correlation between the estimated proportion of pyroxenite in the source and $T_P$ **(Fig. 8a)** or LAB thickness (also $P_f$, **Fig. 8b**). This result contrasts with Sobolev et al.'s (2007) conclusions suggesting that more energetic (i.e., hotter) plumes were more likely to carry a large amount of recycled crust into the upper mantle due to higher buoyancy. However, it is worth noting that the model of Herzberg and Asimow (2015) assumes a peridotite source. Herzberg and Asimow (2008) pointed out that if the mantle source contains pyroxenite and/or volatiles, it might result in overestimation of the calculated potential temperatures by ~100°C. This is consistent with Mallik and Dasgupta (2014) who showed that an OIB source with 120-1830 ppm $CO_2$ (Dasgupta and Hirschmann, 2010) translate to an excess estimation of $T_P$ between 2 and 73 °C (assuming 5-10% partial melting at the source and incompatible behavior of $CO_2$ during partial melting). This calls for a recalibration of mantle thermometers in the future, to take into account the effects of a heterogeneous and volatile-bearing mantle. In fact, a higher $T_P$ will result in a lower contribution of the pyroxenite component in the magma (**Fig. 3**). In other words, overestimating the potential temperature will also result in an overestimation of the pyroxenite fraction required in the source for a given proportion of pyroxenite in the magma. Nevertheless, consistent estimates for the potential temperatures for Hawaii and MOR by other independent methods, such as calibrating major element compositions of liquids with temperature (Lee et al., 2009; Purtika, 2008), suggest that this temperature overestimation is limited and support the decoupling between the thermal state of the lithosphere and the amount of recycled material in the mantle.

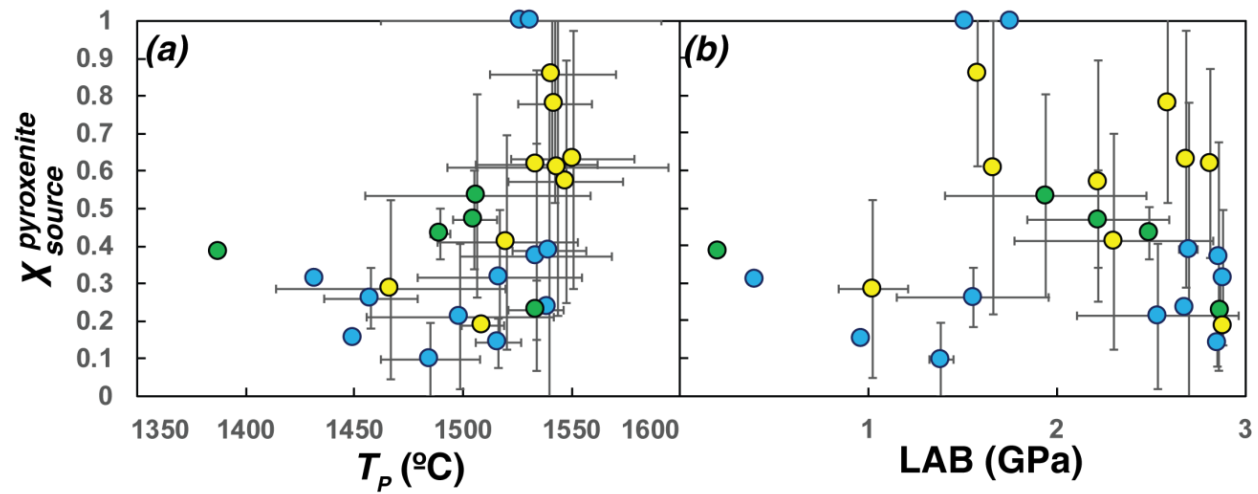

**Figure 8.** Calculated average proportions of G2 using the simple melting model (first set of calculations) as a function of the average $T_P$ (a) and the average pressure of the LAB (b). Same color code as in **Fig. 7**. The error bars correspond to the standard deviation on each island group calculated from fractions obtained for individual island.

*4.2.2 Natural OIB & their cumulates*

OIB display a wider range of major element compositions compared to MORB, but not nearly as wide and variable as arc lavas **(Fig. 6).** Both MORB and OIB show a population density peak around Mg# = 55, suggesting similar degrees of crystal fractionation. Thus, we could infer that the fractionation process for MORB and OIB are relatively similar and the larger range of compositions displayed by OIB is not due to fractionation processes, but instead reflect a larger variability in primary magma compositions. In detail, major element trends of OIBs differ from MORB trends in the following ways: 1) OIB compositions extend to lower $SiO_2$ over similar Mg# intervals compared to MORB, 2) OIBs have higher $TiO_2$ (>2.5 wt.%) than MORB (~2 wt.%), and much higher than arc lavas (<2 wt.%) , 3) Fe-enrichment is not as extreme in OIB as it is for MORB, 4) OIBs follow a trend of $Al_2O_3$ enrichment with decreasing Mg#, whereas MORB become depleted in $Al_2O_3$ , 5) OIBs achieve higher and more variable $Na_2O$ and $K_2O$ contents compared to MORBs **(Fig. 6, Supplementary Fig. S1)**. The spread in $FeO_T$, $Al_2O_3$, $Na_2O$, and $K_2O$ in OIBs versus MORBs likely reflects larger variability in primary OIB magma compositions. Part of this contrast of variability between OIB and MORB is due to the extensive magma mixing at mid-ocean ridges that significantly restricts the range of primitive compositions displayed by MORB (Batiza, 1984; Cipriani et al., 2004). In all the binary oxide plots **(Fig. 6, middle panel)**, the contoured OIB data are skewed toward higher Mg# values compared to MORB, and (near-) primary magmas are commonly sampled, suggesting more limited mixing in oceanic island settings. In addition, and as previously noted, pyroxenites and peridotites produce melt with very similar major element compositions at pressures relevant for mid-ocean ridges (i.e., 1-1.5 GPa, Lambart et al., 2009; Borghini et al., 2017). However, at higher pressures, pyroxenite melting relationships are controlled by the presence of the pyroxene-garnet thermal divide (Kogiso et al., 2004) and can produce highly contrasted melt compositions. Hence, the same amount of lithological heterogeneity in the mantle source results in a higher diversity of magma compositions in OIB settings than in MORB settings. The silica-poor OIB compositions, for instance, are consistent with the partial melting of Si-deficient pyroxenites at high pressure (Hirschmann et al., 2003). As previously mentioned, the high $TiO_2$ contents are also consistent with the presence of recycled oceanic crust in the OIB source.

Finally, the sources of OIBs are usually assumed to be more heterogeneous than the mantle source of MORBs, mostly because primary OIBs also display larger variability in trace element and isotopic ratios than MORBs (Hofmann, 1997, 2003). Two causes for OIBs being more heterogeneous in their source than MORBs are proposed as follows. OIBs show a wider range of $T_P$, spanning to higher $T_P$s than that of MORBs **(Fig. 5a),** and a hotter mantle may result in increased thermal buoyancy of dense garnet-bearing heterogeneities, and therefore, easier for such heterogeneities to accumulate within the upper mantle source of the OIBs (Brown and Lesher, 2014; Sobolev et al., 2005). Also, a thicker lithosphere beneath OIBs, and therefore a higher final pressure of melting, may preserve the chemical signature of heterogeneities better due to lesser degree of partial melting of peridotites (Sobolev et al., 2007). However, we do not observe any correlation between the proportion of chemical heterogeneity and $T_P$ or $P_f$ **(Fig. 8)**, implying that the thermal state of the mantle and the proportion of mantle heterogeneity are decoupled. This casts doubt on the above two propositions about why MORBs may have less heterogeneous sources than OIBs. Ultimately the answer may lie in the process of melt generation such that MORBs undergo greater magma mixing than OIBs (e.g. Zindler et al., 1984), resulting in a tighter range in the concentrations of trace elements and isotopic ratios, and the presence of variability in geochemical signatures in OIBs (or their lack in MORBs) may not be an indicator of how heterogenous their sources are.

4.3. Arc magmas

*4.3.1 Arc magma genesis - a complexity of factors*

Compared to the relatively simple geometry of spreading centers at mid-ocean ridges and the non-tectonic origin of OIBs, the geodynamic setting of volcanic arcs is far more complex, involving multiple moving parts: a hydrated downgoing plate, a convective mantle wedge continuously fluxed by volatiles from the downgoing plate, and overriding plate lithosphere, which itself may be old and heterogeneous (e.g. at continental arcs). As a result, arc magmas worldwide show incredible compositional diversity compared to MORBs and OIBs, reflecting the interplay of the different tectonic components involved. Arc magmas range from basalts (40 – 52 wt.% $SiO_2$) to rhyolites (>70 wt.% $SiO_2$) **(Fig. 6i).** Mg#'s of arc volcanic rocks range from <10 to ~80 **(Fig. 6, right panel).** Such compositional variation is attributed to both primary processes – e.g. those involving melting and melt-rock interaction of peridotitic source rocks in the mantle wedge (and generation of primary melts *sensu stricto*), and secondary processes – such as fractional crystallization and assimilation of pre-existing crust. In addition, there is increasing recognition that partial melting and involvement of non-peridotitic source rocks, such as the downgoing, metamorphosed oceanic slab and sediments, may also contribute to arc magma genesis (as discussed in section 4.3.3). Below, we review experimental constraints on primary melt generation in arcs, discuss recent advances in non-peridotitic contributions to arc magmatism, and summarize key major element compositional trends in natural arc magmas and their cumulates.

*4.3.2 Experimental perspectives on primary arc magma generation: melting of the peridotitic mantle wedge*

We will first investigate the formation of primary arc magmas derived from the peridotitic mantle wedge. The geodynamic scenario for such melting, which constitutes the majority of arc magmatism, is devolatilization of the downgoing slab which induces hydrous flux melting of the dominantly peridotitic mantle wedge (Grove et al., 2006; Till et al., 2012). The chemistry of primary arc magmas depends on the pressure-temperature conditions of partial melting in the mantle, composition of sub-arc peridotite and slab fluxes, and the $H_2O$ content, and oxygen fugacity at the source of melting. Previous experimental studies have investigated the effect of $H_2O$ on partial melting of peridotite (e.g., Gaetani and Grove, 1998; Grove et al., 2006; Hirose, 1997; Hirose and Kawamoto, 1995; Tenner et al., 2012; Till et al., 2012), and mixtures of hydrous siliceous crustal melt and peridotite (Mallik et al., 2015, 2016; Pirard and Hermann, 2014, 2015; Prouteau et al., 2001; Rapp et al., 1999). The partial melts span a wide range in total alkali-silica space covering the entire spectrum of primary arc magmas from basaltic to andesitic and basaltic trachy-andesitic **(Fig. 9a).**

Given the importance of $H_2O$ in subduction zone magmatism, it is important to constrain the effect that $H_2O$ has on the chemistry of magmas produced in the sub-arc mantle and experiments investigating partial melting behavior of the sub-arc mantle can give us insight into this. Experimental melt compositions that co-exist with both olivine and orthopyroxene (the two most abundant minerals in the upper mantle) are compiled and their $H_2O$ contents are plotted against their silica contents **(Fig. 9b).** Two trends emerge. Melt compositions produced at pressures lower than 2 GPa show increasing silica contents with $H_2O$, whereas, for melt compositions produced at pressures greater than 2 GPa, silica contents are anti-correlated with $H_2O$ **(Fig. 9b).** This observation indicates that pressure and $H_2O$ contents affect the chemistry of partial melts, and approach towards the development of a model of subduction zone melting requires that we understand the reason behind such a trend. Mallik et al. (2016) explained that the increasing $SiO_2$ trend at lower pressures is caused by the expansion of olivine stability over orthopyroxene with $H_2O$. The reverse trend at higher pressures is caused by an expanded orthopyroxene stability field over olivine. Thus, future development of a holistic model of partial melting of the mantle in subduction zones requires in-depth understanding of how phase equilibria are affected by such factors . While the effect of pressure, temperature and bulk composition (including $H_2O$ content) have been investigated in previous studies, the effect of oxygen fugacity on differentiation in arcs is still debated.  It has long been known that arc lavas are more oxidized than MORBs (Carmichael, 1991), but the origin of the oxidized signature remains debated. Analyses of $Fe^{3+}$/total Fe in primitive melt inclusions from subduction zones also indicate that arc lavas are more oxidized than mid-ocean ridge basalts and back-arcs, suggesting an origin from a mantle wedge that has been oxidized by slab-derived materials (Brounce et al., 2014, 2015; Grocke et al., 2016; Kelley and Cottrell, 2009; Tollan and Hermann, 2019). However, other studies using redox-sensitive trace elements (Lee et al., 2005; Lee et al., 2010) and isotopic constraints (Dauphas et al., 2009) challenge this hypothesis , and suggest instead that arc lavas acquired their oxidized state via fractional crystallization (Chin et al., 2018; DeBari and Greene, 2011; Jagoutz, 2010; Lee et al., 2006; Tang et al., 2018, 2019), mixing and crustal assimilation (Grove et al., 1982; Patiño Douce,1999), or upper crustal differentiation and open system processes (Blatter et al., 2013; Humphreys et al., 2015).

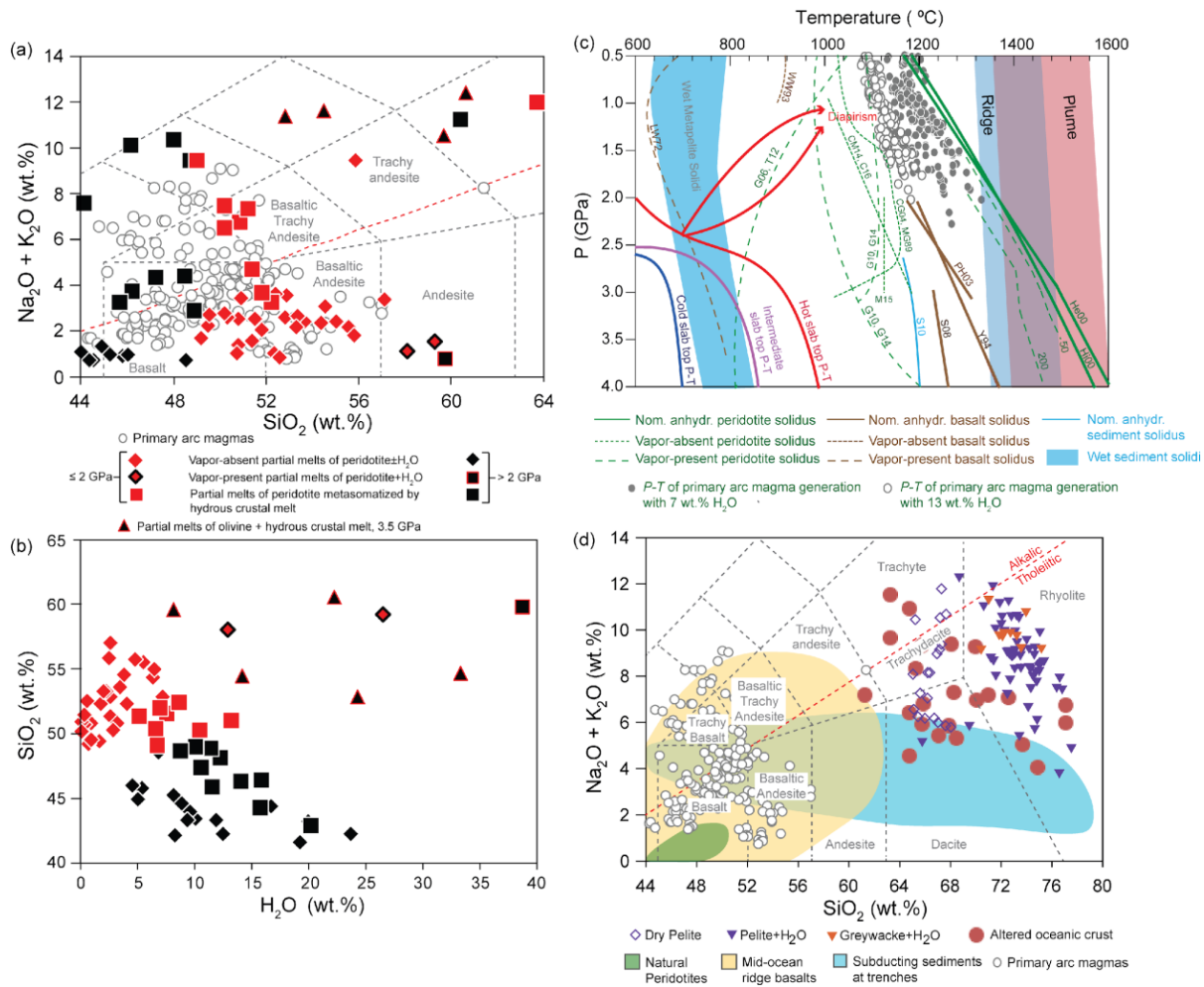

**Figure 9.** (a) Experimentally produced peridotite partial melts without $H_2O$, with $H_2O$ but no co-existing aqueous vapor, and co-existing with aqueous vapor plotted in $SiO_2$ versus total alkali space . Partial melts of peridotite metasomatized by hydrous crustal partial melts are also plotted . (b) $H_2O$ versus $SiO_2$ concentrations of experimentally derived partial melts in equilibrium with olivine and orthopyroxene. This subplot is a modification of **Figure 7** in Mallik et al. (2016). All compositions plotted in the subplots have been normalized to a volatile-free basis. (c) P-T space showing the solidi of nominally anhydrous peridotite, peridotite with 50 and 200 ppm $H_2O$, wet but vapor absent peridotite solidi, nominally anhydrous oceanic crust or basalt, vapor present basalt, wet but vapor absent basalt, nominally anhydrous sediments and wet sediments. The ridge and plume adiabats as well as the hot, intermediate and cold subduction geotherms are the same as in **Figure 5**. Primary arc magma compositions with 7 and 13 wt.% $H_2O$, corrected to be in equilibrium with olivine of Mg# 91 are also plotted. (d) $SiO_2$ versus total alkali ($Na_2O + K_2O$) space with partial melts of sediments and altered oceanic crust produced in experiments, primary arc magmas (same as plotted in panel a), and naturally occurring peridotites, mid-ocean ridge basalts and subducting sediments (are plotted. The classifications of rock types in the figure are based on Le Bas et al., (1986). For details about primary arc magma calculation and he references for the sources of data, see Supplementary Text S1.

*4.3.3 Melting of non-peridotitic sources in arc magma generation.*

*Downgoing plate contributions to fluid flux & sediment melt diapirs.* Based on the similarity of geochemical signatures between subducting sediment and volcanic arc lavas (Plank, 2005), and reproduction of the geochemistry of arc basalts by dehydration of subducted slab (sediments and hydrothermally altered oceanic crust) along with subsequent metasomatism of the sub-arc mantle (Tatsumi and Kogiso, 2003), the involvement of the subducted slab in arc magma production is well-established. Aqueous fluids released from the dehydration of subducted sediments and altered oceanic crust lower the solidus temperatures of sub-arc mantle peridotite due to cryoscopic depression of melting temperatures by $H_2O$ flux. Also, these aqueous fluids metasomatize or alter the mineralogy of the sub-arc mantle to produce hydrous minerals such as mica (phlogopite), amphibole and chlorite. The solidus of such metasomatized peridotites are also lower than that of nominally anhydrous peridotite **(Fig. 9c)**. The latest global thermal models of subduction (Syracuse et al., 2010) show that the solidi of fluid-fluxed peridotite and metasomatized peridotite lie in the thermal regime of the mantle wedge, hence, the sub-arc mantle undergoes fluid-fluxed partial melting. However, it is interesting that fluxing fluids from the dehydration of the slab also lower the solidus of the sediments and altered oceanic crust **(Fig. 9c)**. The latest global thermal models of subduction show that for intermediate subduction zones (such as Guatemala/El Salvador) and hot subduction zones (such as Central Cascadia), the aqueous fluid-fluxed solidi of sediments and altered oceanic crust intersect the pressure-temperature paths of the slabs **(Fig. 9c)**. This implies that sediments and altered oceanic crust partially melt in intermediate to hot subduction zones, and the partial melts are expected to participate in primary arc magma formation. Based on geochemical proxies such as elevated Th/Yb and low Lu/Hf, contribution of slab partial melts to the source of arc magma have been inferred for Central Mexican Volcanic Belt (Cai et al., 2014), Sunda and Lesser Antilles (Woodhead et al., 2001), and the Marianas (Tamura et al., 2014) .

The following two observations can be made about partial melt of subducted crust and the dynamics of subduction zones. Firstly, the partial melts of subducted sediments and altered oceanic crust are much more siliceous ($SiO_2$ > 60 wt.%) than primary arc basalts **(Fig. 9d),** hence, by themselves cannot explain the chemistry of primary magmas produced in subduction zones. Also, the partial melts are more siliceous than partial melts derived from peridotites **(Fig. 9d).** Secondly, according to the thermal models of subduction zones, the isotherms are very closely spaced along the slab top, thus, the slab-top can be as much as 400 °C colder than the overlying sub-arc mantle within a narrow distance of less than around 25 km (Syracuse et al., 2010). These observations imply that slab-derived partial melts are colder (i.e. in thermal disequilibrium) and also chemically distinct (i.e. in chemical disequilibrium) from the adjacent sub-arc mantle. Therefore, as described before, slab partial melts undergo reactive crystallization as they react with the overlying sub-arc mantle (Johnston and Wyllie, 1989; Mallik et al., 2015, 2016; Prouteau et al., 2001; Rapp et al., 1999; Sekine and Wyllie, 1982). Previous studies have shown that such melt-rock reaction produces geochemical signatures of certain primitive arc magma types such as high Mg-andesites and dacites (Prouteau et al., 2001; Tatsumi, 2001; Kepezhinskas et al., 1996), adakites (Rapp et al., 1999) and potassic to ultra-potassic arc magmas (Mallik et al., 2015, 2016).

Geodynamics of subduction zones are further complicated by the potential for buoyant diapirs rising from the subducted slab. As shown by previous studies from geodynamic simulations and geochemistry (Behn et al., 2011; Castro and Gerya, 2008; Marschall and Schumacher, 2012; Nielsen and Marschall, 2017), a mélange comprised of physical mixtures of sediment, altered oceanic crust and metasomatized slab lithosphere can advectively ascend into the overlying hotter mantle wedge from the subducted slab, where it can partially melt and contribute to primary arc magma formation **(Fig. 9c).** Partial melting of physical mixtures of mélange-like compositions and metasomatized lithosphere have been shown to produce alkaline arc magmas (Cruz-Uribe et al., 2018) and ultrapotassic to potassic magmas in arcs and cratons (Förster et al., 2017, 2018; Wang et al., 2017). Finally, the hybridisation of the mantle wedge by melts derived from mélange can also produce a range of primary arc magmas, from tholeiitic to calc-alkaline composition (Codillo et al., 2018).

*Upper plate processes*

In addition to contributions from the downgoing slab, arc magmas may also be influenced by processes in the upper (overriding) plate lithosphere. Since the seminal work of Hildreth and Moorbath (1988) in the Chilean Andes, the concept of a "MASH" (mixing, assimilation, storage, homogenization) zone in the overriding plate has become widely accepted and refined by experimental petrology and geodynamic modeling. The most recent iteration of the MASH hypothesis, Annen et al.'s (2006) Deep Crustal Hot Zone model involves both fractional crystallization of primary arc magmas in the uppermost mantle and deep crust, as well as partial melting of pre-existing crustal rocks in the upper plate. Although by no means an exhaustive list, detailed field studies of arc plutonic rocks from Kohistan (Jagoutz, 2010), Famatinian arc, Argentina (Otamendi et al., 2012; Walker Jr et al., 2015), as well as studies of deep crustal cumulate xenoliths from the Sierra Nevada, California (Ducea, 2002; Lee et al., 2006), Arizona (Erdman et al., 2016) support a primary role for crystallization-differentiation as the mechanism of growing new arc crust, and support the petrological requirement that mafic to ultramafic cumulates be the necessary complements to silicic, calc-alkaline batholiths that are the hallmark of convergent margins. Crystal fractionation of so-called "primary" arc magmas may occur even deeper than the lower crust, as evidenced by residual arc mantle lithosphere that experienced refertilization by basaltic arc melts, resulting in the precipitation of clinopyroxene and garnet ("arclogites") as the melts evolved at depth (Chin et al., 2014; Chin et al., 2016). However, similar pyroxenite enrichment can also occur via melt-rock reaction in the convecting mantle wedge (Berly et al., 2006; Green et al., 2004; Kelemen et al., 1998), and recent studies of global arc volcano trace element systematics suggest that the mantle wedge itself may be periodically metasomatized by enriched continental lithosphere that is eroded during arc magmatism (Turner and Langmuir, 2015). It is thus difficult to separate the contributions from a "contaminated" wedge vs. lithospheric components by using arc volcanic rocks and shallow plutons alone, since both experienced protracted evolution throughout the arc crust, and therefore likely record a combination of both processes (Farmer et al., 2013). Nevertheless, mass balance constraints for the bulk continental crust require at least a 2:1 (but likely to be 10:1 or more) ratio of mafic-ultramafic cumulates and restites as counterparts to the silica-rich arc/continental crust; provided those cumulates are not always convectively removed due to negative buoyancy, they make up the greatest contribution to the overriding arc crust. This, combined with the long-lived (but episodic) nature of continental arcs (Cao et al., 2017; Ducea et al., 2015), indicates that magmatic differentiation and the fractionation of massive amounts of mafic cumulates must be a primary driver of the compositional evolution of arc magmas. Thus, in addition to assimilation and partial melting of pre-existing (and usually older) crustal rocks of the upper plate lithosphere, partial melting of upper plate cumulates may also contribute to the wide variation in arc magma compositions. Foundering, eclogitized lower crust may heat up and partially melt as it descends into the mantle (Gao et al., 2004; Lustrino, 2005), providing yet another way to generate buoyant, silica-rich melts that rise up and become incorporated into the continental crust. As discussed in an earlier section, the solidus of pyroxenitic assemblages is lower than that of peridotitic ones, and so pyroxenites are more easily melted. Experimental partial melts of clinopyroxene-rich cumulates similar to those found in arcs (Médard et al., 2005) show that their partial melts are Ca-rich and Si-undersaturated and similar to some primitive arc melt inclusions (Schiano et al., 2000).

*4.3.4 Natural arc magmas & their cumulates*

Erupted primary lavas in volcanic arcs are rare, as they are usually modified by secondary processes as the melt migrates from the mantle source to eruption at the surface. Direct samples of putative primary arc melts occur as melt inclusions hosted in high Mg# olivine phenocrysts. Such high Mg# (89 – 92) olivines are in equilibrium with mantle peridotite, and thus it is assumed that any liquid trapped in such olivines are also mantle-derived and therefore primary. However, olivine-hosted melt inclusions are not sampled in every arc, are not as abundant as erupted primitive lavas, and also may experience diffusive re-equilibration, potentially compromising any primary chemical signatures (Gaetani et al., 2012). In the right hand panels of **Figure 6**, we plot nearly 40,000 compositions of arc volcanic rocks downloaded from the GEOROC online database, nearly 400 cumulates of arc magmas previously compiled by Chin et al. (2018), and 205 primary arc melt inclusion compositions (see **Supplementary Table S5** for references). One of the clearest features of the data to emerge from a simple visual inspection of the major element plots is that, unlike the tightly coherent trends described by MORBs and their cumulates, arc volcanic rocks show considerable scatter and a much wider range in nearly every major element. By contrast, melt inclusions plot in a restricted Mg# window but interestingly span a wide range of compositions in certain elements, notably CaO, $K_2O$, and $Na_2O$ **(Figs. 6, S1)**. Such large variations at high Mg# suggests primary melts sample heterogeneous sources in the mantle wedge and overriding plate (Sadofsky et al. 2008). Recent studies on arc lavas using trace element systematics also point to an important role of mantle wedge heterogeneity in generating the diversity of arc magma compositions (Turner and Langmuir, 2015; Turner et al., 2016). The utility of major elements in fingerprinting subduction fluxes is limited compared to the sensitivities of trace element compositions, trace element ratios, and isotopic data, but a comprehensive survey of all the available data is beyond the scope of this contribution.

As seen in the right hand panels of **Figure 6**, arc cumulates also generally span a larger compositional variation than MORB and OIB cumulates. This observation implies that arc magmas and their cumulates reflect more complex petrogenesis compared to MORBs and OIBs, as discussed earlier. For example, despite the highest concentration of arc lavas falling within a similar range of $SiO_2$ content as MORB and OIB (darkest red contours in **Fig. 6i**), the contoured arc lava data skew towards higher $SiO_2$, in contrast to MORB and OIB, the former having a more or less symmetrical distribution in $SiO_2$ and the latter skewing towards lower $SiO_2$. One explanation for the overall higher $SiO_2$ in arc lavas is that they experience a greater degree of fractional crystallization and/or assimilation of crustal rocks, starting at fairly primitive compositions and therefore deep and early in their evolution (Lackey et al., 2005; Nelson et al., 2013).

When arc volcanic rocks and arc cumulates are viewed together, several complementary trends consistent with an important role for fractional crystallization emerge. In the contoured data of CaO vs. Mg#, most arc volcanic rocks have CaO <10 wt.%, but MORBs have CaO >10 wt.% **(Supplementary Fig S1a,g)**. This can be explained by the early precipitation of anorthite and/or clinopyroxene as a major cumulate mineral in arcs, whereas in mid-ocean ridges plagioclase saturates early and tends to be albitic (Beard, 1986; Chin et al., 2018; Grove et al., 1993). This is also borne out in the systematics of $Al_2O_3$ and $Na_2O$ **(Fig. 6j,k)**. MORB cumulates define a tight trend in $Al_2O_3$ vs. Mg# space and show a peak in $Al_2O_3$ at high Mg# **(Fig. 6b),** whereas arc cumulates define a looser trend with a weaker peak at high Mg# **(Fig. 6j).** Indeed, the median $Al_2O_3$ of the most primitive MORB cumulates (Mg# 80 – 90) is substantially higher than arc cumulates in the same Mg# range (Chin et al. 2018 and **Figure 3** therein). These cumulate systematics subsequently control the $Al_2O_3$ content of differentiating arc magmas. In MORBs, $Al_2O_3$ defines a tight, decreasing trend as Mg# decreases, reflecting the simple mineralogy of MORB cumulates (gabbroic). By contrast, $Al_2O_3$ in arc volcanic rocks span a wide range of $Al_2O_3$ (10 – 20 wt.%) with decreasing Mg#, indicating multiple mineral contributions (pyroxene, amphibole, etc.) to the fractionating cumulate assemblage in contrast to the simple gabbroic mineralogy of MORB cumulates, including an important role for garnet (Tang et al. 2018), especially in continental arcs. Most arc lavas have $Al_2O_3$ between 16 – 18 wt.% (e.g., the classic "high alumina arc basalts", (Crawford et al., 1987)) compared to MORBs, which peak below 16 wt.% $Al_2O_3$, and, importantly, skew towards lower $Al_2O_3$ compared to arc lavas **(Fig. 6j).** The plot of $Na_2O$ vs. Mg# also reflects the plagioclase-dominated gabbroic cumulates of MORB compared to arcs and complementary trends in the lavas. Lastly, as discussed in the MORB section, the classic tholeiitic vs. calc-alkaline trends clearly emerge in comparing MORBs vs. arcs. The diagnostic feature of calc-alkaline lavas is Fe-depletion with increasing differentiation **(Ssupplementary Fig. S1i)**, attributed to crystal fractionation of Fe- and Ti-bearing phases such as garnet and magnetite **(Fig. 6l)**.

## 5. Future directions

The summary of experimental data presented in Section 2 and in previous review articles (Kogiso et al., 2004; Lambart et al., 2013) emphasize the large range of compositions covered by the potential lithologies present in the mantle source of magmas. This makes it challenging to determine a unique proxy as a tracer of heterogeneities in the mantle source of magmas. In addition, calculations presented in Section 4.2 highlight that different proxies can result in different estimates of the nature and proportion of the heterogeneity in the mantle. Similarly, compositional observations of natural MORB and their cumulates point out a decoupling between the major element compositions that support fractional crystallisation as the main process of magma differentiation, and trace element compositions that indicate the strong influence of porous flow and melt-rock interactions. Hence, it is crucial for the geochemical community to concentrate their future efforts into the construction of a comprehensive dataset, with studies combining major-element, trace element and isotopic analyses. In order to correctly interpret the geochemical dataset, we need parameterizations and geochemical models that take into account such a large range of compositions. Hence, we need better constraints on the role of the lithological composition on partial melting behavior and element partitioning, especially in the presence of volatiles. As far as volatiles are concerned, the focus has largely been on $H_2O$ and/or $CO_2$, but a growing consensus has emerged that other species, notably S and N are important contributors to the solid Earth volatile recycling budget (Evans, 2012; Cartigny and Marty, 2013; Mallik et al., 2018), and so also influence the partial melting behavior and consequent differentiation of the upper mantle. Future studies should not only investigate the effect of S and N, but should also constrain the effect of the presence of mixed volatile species of H-C-N-S on partial melting of the upper mantle.

As mentioned earlier in this chapter, the thermal state of the mantle, expressed in terms of potential temperature ($T_P$) is an extremely important aspect in the chemical differentiation of the Earth. $T_P$ is currently estimated based on the magnesian content of erupted primary lavas, assuming that a hotter mantle produces magmas with higher magnesia contents (e.g. Herzberg et al., 2007; Herzberg and Gazel, 2009). It has been shown that melt-rock reactions involving heterogeneities in the upper mantle and the presence of volatiles produce magnesia-rich magmas at colder mantles than that predicted earlier (Mallik and Dasgupta, 2014). This begs for further studies that would calibrate a mantle thermometer to estimate $T_{PS}$ taking into account factors such as melt-rock reactions and influence of volatiles, else we may overpredict the thermal state of the Earth's mantle beneath tectonic settings.

As far as subduction zones are concerned, there is broad agreement on the behavior of aqueous fluids and partial melts from the subducting slab and their effects on generation of arc magmas, and transfer of elements from the subducted crust to the mantle. However, one of the least understood phases in subduction zones are supercritical fluids, which are the products of complete miscibility between aqueous fluids and partial melts. The reason why supercritical fluids remain so poorly understand is due primarily to experimental difficulty in studying them. Most experimental techniques employ ex-situ analyses of samples, where the sample, after being "cooked" at the P-T of interest, are quenched to room P-T and investigated, assuming that modifications during the quenching process are not significant. Kessel et al. (2004) went a step further and cryogenically preserved the fluid in their quenched capsules during laser ablation mass spectroscopy, in order to preserve the fluid phase in their capsules. However, supercritical fluids may separate into two phases during the quenching, rendering ex-situ analyses of this phase nearly impossible. A few studies have experimentally demonstrated the closing of the miscibility gap between aqueous fluids and silicate melts under P-T conditions applicable to subduction zones (Bureau and Keppler, 1999; Kawamoto et al., 2012; Mibe et al., 2007, 2011; Ni et al., 2017; Shen and Keppler, 1997) using a Bassett-type hydrothermal diamond anvil cell or X-ray radiography using a multianvil apparatus. However, the current state-of-the-art in such in-situ experimental techniques have limitations in terms of chemical analyses that can be performed to constrain how these fluids may affect element transport in subduction zones. A future challenge for experimental geoscientists would be to improve the techniques of in-situ analysis such that supercritical fluids can be investigated better.

Finally, melt-rock interactions, in the mantle and in the crust, are increasingly recognized as important petrogenetic processes, both in the mantle and in the crust, across all geological settings. In our view, one of the major challenges faced by the experimental community is to develop tools and provide experimental constraints to predict the system behavior during these interactions including complex feedback between composition and melting and crystallization behavior and kinetics of interactions during melt transport. Some of the questions we need to address are: (1) the required conditions for the preservation of the source signal in primary magmas, (2) the effect of chromatographic melting on the range of isotopic compositions and major and trace elements concentrations in MORB and OIB (Navon and Stolper, 1987) , (3) the mechanisms of transition between the various types of magma flow in the upper mantle (porous versus focused flow; e.g., Kelemen et al., 1997), and (4) the implications of crustal melt-rock interaction on the composition of the cumulate minerals and lavas (e.g., Lissenberg et al., 2013).

**Acknowledgments, Samples, and Data**
AM thanks the editors for their invitation to contribute to this special volume and Katie Kelley for discussions about sources of primary arc melt inclusion data. EJC thanks GM Bybee, K

Shimizu, and CTA Lee for informal discussions regarding deep crustal arc cumulates. SL was supported by the NSF grant EAR-1834367. We also thank the reviewers for their constructive comments, especially the thorough review from Fred Davis which really helped improve the manuscript. The data supporting the findings of this study are available in the Figshare data repository: Text S1, Fig. S1 and Table S5: https://doi.org/10.6084/m9.figshare.9926867.v1 ; Tables S1 to S4 and S6 to S9: https://doi.org/10.6084/m9.figshare.9926885.v1